\documentclass[english,aps,twocolumn]{revtex4-1}
\usepackage[T1]{fontenc}
\usepackage[latin9]{inputenc}
\usepackage{amsmath}
\usepackage{amssymb}

\makeatletter
 
 \@ifundefined{textcolor}{}
 {%
   \definecolor{BLACK}{gray}{0}
   \definecolor{WHITE}{gray}{1}
   \definecolor{RED}{rgb}{1,0,0}
   \definecolor{GREEN}{rgb}{0,1,0}
   \definecolor{BLUE}{rgb}{0,0,1}
   \definecolor{CYAN}{cmyk}{1,0,0,0}
   \definecolor{MAGENTA}{cmyk}{0,1,0,0}
   \definecolor{YELLOW}{cmyk}{0,0,1,0}
 }

\makeatother

\usepackage{babel}

\begin{document}

\preprint{This line only printed with preprint option}

\title{Technical aspects of the evaluation of the overlap of Hartree- Fock-
Bogoliubov wave functions}

\author{L.M. Robledo}

\email{luis.robledo@uam.es}

\affiliation{Departamento Física Teórica, Universidad Autónoma de Madrid, E-28049
Madrid, Spain}
\begin{abstract}
Several technical aspects concerning the evaluation of the overlap
between two mean field wave functions of the Hartree Fock Bogoliubov
type, are discussed. The limit when several orbitals become fully
occupied is derived as well as the formula to reduce the dimensionality
of the problem when exactly empty orbitals are present. The formalism
is also extended to deal with the case where the bases of each of
the wave functions are different. Several practical results concerning
the evaluation of pfaffians as well as the canonical decomposition
of norm overlaps are also discussed in the appendices.
\end{abstract}
\maketitle

\section{introduction}

In a recent publication \cite{Robledo.09} a new formulation, based
on the pfaffian of a skew-symmetric matrix, has been proposed to compute
the overlap between mean field wave functions of the Hartree Fock
Bogoliubov (HFB) type, including its phase (or sign, for real overlaps).
The result for the overlap is obtained by recursing to the powerful
concept of fermion coherent states \cite{Klauder.85,Berezin.66,Blaizot.85,Negele.88,Ohnuki.78}
and it involves the evaluation of a quantity called the pfaffian of
a skew-symmetric matrix -see, for instance \cite{Caianello} for a
definition of the pfaffian in a physical context- that is similar
in spirit (linear combinations of products of matrix elements) to
the determinant of a general matrix. In the derivation of the formula
for the overlap \cite{Robledo.09} it is assumed that the two mean
field wave functions (of the HFB type -see \cite{RS.80} for definition
and properties-) can be related to a common reference one (usually
chosen as the true particle vacuum) by means of their Thouless parametrization.
It is very likely to find cases where the Thouless parametrization
is ill defined because, for that particular case, it involves the
inverse of a (near) singular matrix. This situation corresponds to
the presence of particles (or quasiparticles in the general case)
that have an occupancy of one, rendering the wave function orthogonal
to the common (or reference) wave function. To handle those singular
cases it was suggested in \cite{Robledo.09} to just change the common
reference mean field wave function in order to modify the occupancies
with the hope that none of them will be close to one. However, it
is desirable to have an alternative for those cases where the change
of reference wave function is either not possible or too cumbersome
to carry out. Therefore, I have considered the formal limit of occupancies
going to one and I have obtained a formula which is well defined in
that limit and provides a sound answer to such singular limit. By
using the same kind of ideas I have also handled explicitly the case
where some particles (or quasiparticles) have zero occupancy and therefore
they do not contribute to the overlap. The formula obtained in this
case involves matrices of smaller dimensions than the original ones
and therefore should be regarded as a thrifty alternative to the original
formula for situations where the wave functions are expanded in huge
bases. Similar manipulations to reduce the size of matrices and/or
deal with full occupancies, have been considered in Refs \cite{Bonche.90,Haider.92,Robledo.94,Yao.09}.
I also address the case where each of the mean field wave functions
are expressed in different single particle basis related to each other
by a general (not necessarily unitary) transformation. The result
obtained is useful, for instance, to compute the overlap of the operators
for spatial transformations (as translations or rotations) between
arbitrary mean field HFB wave functions. The result is general enough
as to allow for transformations that do not map the single particle
basis into itself (non complete basis under the transformation). This
was already considered in \cite{Robledo.94} in a general framework
but not considering the present formulation including the phase of
the overlap. In \cite{Haider.92} the implications of considering
two different bases are also address, but there it is implicitly assumed
that both bases share the same block structure defining the conjugate
states. This is a limiting assumption that do not hold in general
for time reversal violating (cranking, for instance) wave functions.
Finally, some useful results concerning the formal evaluation of pfaffians
are discussed in the appendices. These results can be of interest
in other branches of physics where the use of pfaffians is becoming
increasingly popular \cite{Bajdich.08,Thomas.09,Stephan.09}. To cover
also more practical aspects, the reader is referred to \cite{CG-B.11}
for a thorough description of useful algorithms to compute numerically
and symbolically the pfaffian of arbitrary skew symmetric matrices.

The relevance of the results present here is a direct consequence
of the increasing popularity of the so called {}``beyond mean field
methods'' in nuclear physics \cite{Bender.03,Yao.09,rodriguez.02,SPM.07,Schmid.04,Bender.08}
that demand the evaluation of both the modulus and phase of the overlaps
between arbitrary HFB wave functions. A reliable determination of
the sign of the norm can also be useful in to order to pin down the
location of the zeros of the HFB overlaps \cite{Oi.05}. This determination
would eventually be useful to get rid of the so called {}``pole problem''
that plagues present beyond mean field calculations \cite{Donau.98,anguiano.01,Lacroix.09}. 

In section \ref{sec:Limits} the formulas pertaining the two limits
considered are derived and their implications discussed. We also explicitly
show how to implement the change to a different common reference HFB
state that could be an easy alternative in some cases. In section
\ref{sec:DiferentBases}the case where the two HFB wave functions
are referred to different bases is discussed. Finally, in appendices
\ref{sec:Overlap}, \ref{sec:BipartitePfaff}, and\ref{sec:ExchMat}
some relevant results required in the derivations are discussed.

\section{Evaluation of the overlap in various limits\label{sec:Limits}}

To begin with, it is convenient to recall the notation used in \cite{Robledo.09}
as well as the formulas obtained there. The goal is to evaluate both
the modulus and phase of the overlap $\langle\phi_{0}|\phi_{1}\rangle$
between two HFB wave functions $|\phi_{0}\rangle$ and $|\phi_{1}\rangle$
which are given in terms of the Thouless parametrization of a general
HFB wave function

\begin{equation}
|\phi_{i}\rangle=\exp\left(\frac{1}{2}\sum_{kk'}M_{kk'}^{(i)}a_{k}^{+}a_{k'}^{+}\right)|0\rangle\label{eq:HFBWF}\end{equation}
The skew-symmetric matrices $M^{(i)}=(V^{(i)}U^{(i)-1})^{*}$, of
dimension $N\times N$ ($N$ is assumed to be an even number $N=2q$
as required for fermions), are related to the coefficients $U^{(i)}$
and $V^{(i)}$ of the Bogoliubov transformation defining the quasiparticle
annihilation operators 

\[
\alpha_{k}^{(i)}=\sum_{l}U_{lk}^{(i)*}a_{l}+V_{lk}^{(i)*}a_{l}^{+}\]
associated to $|\phi_{i}\rangle$. The above wave functions are not
normalized to one, but as to have $\langle0|\phi_{i}\rangle=1$ instead.
As shown in \cite{Robledo.09} the overlap can be written as \begin{equation}
\langle\phi_{0}|\phi_{1}\rangle=(-1)^{N(N+1)/2}\textrm{pf}\mathbb{M}\label{eq:Over}\end{equation}
with\begin{equation}
\mathbb{M}=\left(\begin{array}{cc}
M^{(1)} & -\mathbb{I}\\
\mathbb{I} & -M^{(0)*}\end{array}\right)\label{eq:MM}\end{equation}
is a $2N\times2N$ matrix. To obtain the results of the present section
the Bloch-Messiah decomposition of the Bogoliubov amplitudes \cite{RS.80},
namely $U^{(i)}=D^{(i)}\bar{U}^{(i)}C^{(i)}$ and $V^{(i)}=D^{(i)*}\bar{V}^{(i)}C^{(i)}$,
is used. In the previous expressions, $D^{(i)}$ and $C^{(i)}$ are
given unitary matrices and $\bar{U}^{(i)}$ and $\bar{V}^{(i)}$ are
real matrices with special diagonal forms. By using this decomposition
we can write \begin{equation}
M^{(i)}=D^{(i)}M_{C}^{(i)}D^{(i)\, T}\label{eq:Mi}\end{equation}
 where the skew-symmetric matrix $\mbox{M}_{C}^{(i)}$ is in {}``skew-symmetric
diagonal'' (or canonical) form\begin{equation}
\mbox{M}_{C}^{(i)}=\left(\begin{array}{cc}
0 & \bar{M}^{(i)}\\
-\bar{M}^{(i)} & 0\end{array}\right)\label{eq:MiC}\end{equation}
The diagonal matrix $\bar{M}^{(i)}$has matrix elements \begin{equation}
\bar{M}_{jk}^{(i)}=\frac{v_{j}^{(i)}}{u_{j}^{(i)}}\delta_{jk}.\label{eq:Mbar}\end{equation}
The extreme values of the ratios $v_{j}^{(i)}/u_{j}^{(i)}$ are infinity
for fully occupied levels ($v^{(i)}=1$) or zero for empty levels
($v^{(i)}=0$). For further developments it is convenient to single
out those values and write\begin{equation}
\bar{M}^{(i)}=\left(\begin{array}{cc}
\bar{N}^{(i)} & 0\\
0 & \bar{O}^{(i)}\end{array}\right)\label{eq:MdNO}\end{equation}
where the diagonal matrix $\bar{O}^{(i)}$contains the $K^{(i)}$
diagonal elements belonging to the extreme values, infinity or zero,
mentioned above. The dimension of this matrix is $K^{(i)}\times K^{(i)}$.
Each of the two limiting cases require different considerations and
hence we will from now on considered them separately.

\subsection{Limit of fully occupied levels\label{sub:Occ}}

In this case, there are $K^{(i)}$ fully occupied levels in each of
the HFB wave functions $|\phi_{i}\rangle$ and the corresponding diagonal
elements of the matrices $\overline{M}^{(i)}$ (the ones corresponding
to $\bar{O}^{(i)}$ in Eq. (\ref{eq:MdNO})) tend to infinity. This
is a serious challenge, as the overlap Eq (\ref{eq:Over}) as well
as the norm of the $|\phi_{i}\rangle$ diverge. The divergence has
to be regularized and singled out of the overlap in order to cancel
it out with the diverging factors coming from the norms of the HFB
wave functions. To this end we write $M_{C}^{(i)}=R^{(i)}M_{CR}^{(i)}R^{(i)T}$
where we have introduced the {}``canonical regularized'' (CR) matrix
\[
M_{CR}^{(i)}=\left(\begin{array}{cc|cc}
 &  & \bar{N}^{(i)} & 0\\
 &  & 0 & \mathbb{I}_{K^{(i)}}\\
\hline -\bar{N}^{(i)} & 0\\
0 & -\mathbb{I}_{K^{(i)}}\end{array}\right)\]
as well as \begin{equation}
R^{(i)}=\left(\begin{array}{cccc}
\mathbb{I}_{N-K^{(i)}} & 0 & 0 & 0\\
0 & \mathbb{I}_{K^{(i)}} & 0 & 0\\
0 & 0 & \mathbb{I}_{N-K^{(i)}} & 0\\
0 & 0 & 0 & \bar{O}^{(i)}\end{array}\right)\label{eq:Ri}\end{equation}
In all the cases $\mathbb{I}_{K^{(i)}}$ represents the unit matrix
of dimension $K^{(i)}$. With the above definitions the matrix $\mathbb{M}$
of Eq (\ref{eq:MM}), which enters the expression of Eq (\ref{eq:Over})
for the overlap, is factorized as\[
\mathbb{M}=\left(\begin{array}{cc}
\tilde{R}^{(1)} & 0\\
0 & \tilde{R}^{(0)*}\end{array}\right)\left(\begin{array}{cc}
M_{CR}^{(1)} & S\\
-S^{T} & -M_{CR}^{(0)*}\end{array}\right)\left(\begin{array}{cc}
\tilde{R}^{(1)T} & 0\\
0 & \tilde{R}^{(0)+}\end{array}\right)\]
 where \[
\tilde{R}^{(i)}=D^{(i)}R^{(i)}\]
 and \begin{equation}
S=\tilde{R}^{(1)\,-1}\left(\tilde{R}^{(0)+}\right)^{-1}.\label{eq:S}\end{equation}
 Using now the property $\textrm{pf}(B^{T}AB)=\det(B)\textrm{pf}(A)$
we obtain\[
\textrm{pf}(\mathbb{M})=\textrm{det}(D^{(1)})\textrm{det}(D^{(0)*})\textrm{det}(R^{(1)})\textrm{det}(R^{(0)*})\textrm{pf}(\tilde{\mathbb{M}})\]
with \[
\tilde{\mathbb{M}}=\left(\begin{array}{cc}
M_{CR}^{(1)} & S\\
-S^{T} & -M_{CR}^{(0)*}\end{array}\right)\]
In the case of fully occupied levels, the diagonal matrices$\bar{O}^{(i)}$
introduced in Eq. (\ref{eq:Ri}) tend to infinity. As a consequence,
the determinants $\textrm{det}(R^{(i)})$ diverge as do some of the
matrix elements of $\tilde{R}^{(i)}$. The later is not a problem
as just the inverse of this matrix is required in Eq. (\ref{eq:S})
\[
\tilde{R}^{(i)\,-1}=R^{(i)-1}D^{(i)\,-1}\]
and $R^{(i)\,-1}$ is a well defined quantity \[
R^{(i)\,-1}=\left(\begin{array}{ccc}
\mathbb{I}_{N} & 0 & 0\\
0 & \mathbb{I}_{N-K^{(i)}} & 0\\
0 & 0 & 0_{K^{(i)}}\end{array}\right)\]
As a consequence of this structure, the matrix $S$ of Eq (\ref{eq:S})
is, in this limit, a matrix where the last $K^{(1)}$ rows and last
$K^{(o)}$ columns are set strictly to zero. We can use this property
together with the special structure of the matrices $M_{CR}^{(i)}$
to reduce the size of the matrices to be considered in the evaluation
of the pfaffian. This will be addressed in the next subsections in
a slightly different context. The only truly diverging quantities,
namely the determinants $\textrm{det}(R^{(i)})$, cancel out when
we compute the normalized overlap $\langle\varphi_{0}|\varphi_{1}\rangle/\sqrt{|\langle\varphi_{0}|\varphi_{0}\rangle|^{2}|\langle\varphi_{1}|\varphi_{1}\rangle|^{2}}$
as, from the previous formulas, $|\langle\varphi_{i}|\varphi_{i}\rangle|^{2}$
is proportional to $|\textrm{det}(R^{(i)})|^{2}$.

\subsection{Limit of fully empty levels\label{sub:empty} }

Another situation often encounter in numerical applications is when
many orbitals have zero occupancies $v^{2}=0$ and therefore their
contribution to the overlap is zero. To disentangle those contributions
and reduce, in this way, the computational cost of the evaluation
of the norm it is convenient to consider the limit of fully empty
levels. This limit has been considered by other authors \cite{Bonche.90,Haider.92,Robledo.94,Yao.09}
in the past but in the more traditional formulation of the overlap
not including the phase. In the limit of fully empty levels I will
show that the evaluation of the overlap can be reduced to considering
matrices with dimension equal to the number of orbitals with non-zero
occupancy. This is an advantage as the number of non-zero occupancy
levels is usually much smaller than the total dimensionality of the
basis used. I will use the same notation as before, but in this case,
the $\bar{O}^{(i)}$ are diagonal matrices whose diagonal matrix elements
are made to tend to zero. The number of diagonal elements, $K^{(i)}$,
represent the number of empty levels in each HFB wave function $|\phi_{i}\rangle$.
In this case it is convenient to reorder the single particle basis
by using unitary similarity transformations $P_{23}$ (see appendix
\ref{sec:ExchMat}) permuting blocks 2 and 3, to write $M_{C}^{(i)}=P_{23}^{(i)}M_{CR}^{(i)}P_{23}^{(i)^{T}}$with\[
M_{CR}^{(i)}=\left(\begin{array}{cc|cc}
0 & \bar{N}^{(i)}\\
-\bar{N}^{(i)} & 0\\
\hline  &  & 0 & \bar{O}^{(i)}\\
 &  & -\bar{O}^{(i)} & 0\end{array}\right)=\left(\begin{array}{cc}
N^{(i)} & 0\\
0 & O^{(i)}\end{array}\right)\]
where the matrix $N^{(i)}$ has dimension $2(N-K^{(i)})\times2(N-K^{(i)})$
and $O^{(i)}$ has dimension $2K^{(i)}\times2K^{(i)}$. The matrix
carrying out the permutation $P_{23}^{(i)}$ depends on the basis
as the size of each block $\bar{N}^{(i)}$ is not necessarily the
same. The impact of this dependence will show up in the evaluation
of the pfaffian where the determinant of $P_{23}^{(i)}$ is required.
Now, the new unitary matrix $D_{R}^{(i)}=D^{(i)}P_{23}^{(i)}$ is
introduced to write \begin{equation}
M^{(i)}=D_{R}^{(i)}M_{CR}^{(i)}D_{R}^{(i)\, T}.\label{eq:MiR}\end{equation}
Now a technical detail related to the way how to handle the situation
where the bipartite structure of $M_{CR}^{(0)}$ is different from
the one of $M_{CR}^{(1)}$ has to be considered. The different bipartite
structure happens when the dimensions $K^{(0)}$ and $K^{(1)}$ differ,
or in other words, that the number of empty levels in each mean field
wave function is different. The optimal situation in terms of simplicity
is that when both $K^{(i)}$ are the same. This is the strategy we
will use in the following by assuming the same bipartite structure
in both $M_{CR}^{(i)}$ and $D_{R}^{(i)}$ with a common dimension
$K_{S}$ which is the smaller of $K^{(0)}$ and $K^{(1)}$. With this
in mind, we endow the matrices $M_{CR}^{(i)}$ with the new bipartite
structure\begin{equation}
M_{CR}^{(i)}=\left(\begin{array}{cc}
N_{C}^{(i)} & 0\\
0 & O^{(i)}\end{array}\right)\label{eq:NcO}\end{equation}
where $N_{C}^{(i)}$ are $2(N-K_{S})\times2(N-K_{S})$ matrices and
the $O^{(i)}$ have the same dimension $2K_{S}\times2K_{S}$. In the
same way, we will use below the same bipartite structure for the matrix
$D_{R}^{(i)}$ \begin{equation}
D_{R}^{(i)}=\left(\begin{array}{cc}
D_{11}^{(i)} & D_{12}^{(i)}\\
D_{21}^{(i)} & D_{22}^{(i)}\end{array}\right)\label{eq:Dbp}\end{equation}
where $D_{11}^{(i)}$ is a square matrix of dimension $2(N-K_{S})\times2(N-K_{S})$,
$D_{12}^{(i)}$ is of dimension $2(N-K_{S})\times2K_{S}$ , $D_{21}^{(i)}$
is of dimension $2K_{S}\times2(N-K_{S})$ and finally $D_{22}^{(i)}$
is of dimension $2K_{S}\times2K_{S}$. If we now use Eqs. (\ref{eq:NcO})
and (\ref{eq:Dbp}) to reconstruct the matrix $M^{(i)}$ of Eq. (\ref{eq:MiR})
we realize that in the $O^{(i)}\rightarrow0$ limit the sub-matrices
$D_{12}^{(i)}$ and $D_{22}^{(i)}$ do not enter the final expression
of $M^{(i)}$. We can use this freedom to chose those {}``arbitrary''
matrices in such a way as to simplify some of the results to be obtained
below. A possible, and convenient, choice is $D_{12}^{(i)}=0$ and
$D_{22}^{(i)}=\mathbb{I}_{2K_{S}}$. The matrix obtained with this
choice will be denoted by $\overline{D}_{R}^{(i)}$ with \begin{equation}
\overline{D}_{R}^{(i)}=\left(\begin{array}{cc}
D_{11}^{(i)} & 0\\
D_{21}^{(i)} & \mathbb{I}_{2K_{S}}\end{array}\right)\label{eq:BarDR}\end{equation}
 and has the nice property of having a simple inverse \textbf{\[
\overline{D}_{R}^{(i)\,-1}=\left(\begin{array}{cc}
D_{11}^{(i)\,-1} & 0\\
-D_{21}^{(i)}D_{11}^{(i)\,-1} & \mathbb{I}_{2K_{S}}\end{array}\right)\]
}involving only the inverse of the matrices $D_{11}^{(i)}$ which
have a moderate dimensionality. This is the property guiding the choice
made, as we are implicitly assuming that $N-K_{S}\ll K_{S}$ and it
has to be kept in mind that the cost of most of the matrix operations
grow as the cubic power of their dimension. 

With all these definitions the matrix $\mathbb{M}$ which enters the
formula for the overlap is written as,\[
\mathbb{M}=\left(\begin{array}{cc}
\overline{D}_{R}^{(1)} & 0\\
0 & \overline{D}_{R}^{(0)*}\end{array}\right)\left(\begin{array}{cc}
M_{CR}^{(1)} & U\\
-U^{T} & -M_{CR}^{(0)*}\end{array}\right)\left(\begin{array}{cc}
\overline{D}_{R}^{(1)\, T} & 0\\
0 & \overline{D}_{R}^{(0)\,+}\end{array}\right)\]
 where \begin{equation}
U=\overline{D}_{R}^{(1)\,-1}\left(\overline{D}_{R}^{(0)\,+}\right)^{-1}\label{eq:DefU}\end{equation}
 (please note that the unitary character of the $D_{R}^{(i)}$ matrices
is lost with the introduction of the $\overline{D}_{R}^{(i)}$ ones).
Using now the properties of the pfaffian we obtain\begin{eqnarray}
\textrm{pf}(\mathbb{M}) & = & \textrm{det}(\overline{D}_{R}^{(1)})\textrm{det}(\overline{D}_{R}^{(0)*})\textrm{pf}(\tilde{\mathbb{M}})\label{eq:pfMempty}\\
 & = & \det(D_{11}^{(1)})\det(D_{11}^{(0)\,*})\textrm{pf}(\tilde{\mathbb{M}})\end{eqnarray}
with \begin{equation}
\tilde{\mathbb{M}}=\left(\begin{array}{cc}
M_{CR}^{(1)} & U\\
-U^{T} & -M_{CR}^{(0)*}\end{array}\right).\label{eq:Mtilda}\end{equation}
Let us now analyze the structure of the block matrix $U$ entering
the definition of $\tilde{\mathbb{M}}$. Using Eq (\ref{eq:DefU})
together with Eq (\ref{eq:BarDR}) allows to write \[
U=\left(\begin{array}{cc}
U_{11} & U_{12}\\
U_{21} & U_{22}\end{array}\right)\]
with\begin{eqnarray}
U_{11} & = & \left(D_{R\,11}^{(0)\,+}D_{R\,11}^{(1)}\right)^{-1}\label{eq:DefU11}\\
U_{12} & = & -U_{11}D_{R\:21}^{(0)\,+}\label{eq:DefU12}\\
U_{21} & = & -D_{R\:21}^{(1)}U_{11}\label{eq:DefU21}\\
U_{22} & = & \mathbb{I}+D_{R\:21}^{(1)}U_{11}D_{R\:21}^{(0)\,+}\label{eq:DefU22}\end{eqnarray}
With this definition, the matrix $\tilde{\mathbb{M}}$ of Eq (\ref{eq:MtildaExt})
acquires, in the limit where $O^{(i)}\rightarrow0$, the block structure\[
\tilde{\mathbb{M}}=\left(\begin{array}{cc|cc}
N_{C}^{(1)} & 0 & U_{11} & U_{12}\\
0 & 0 & U_{21} & U_{22}\\
\hline -U_{11}^{T} & -U_{21}^{T} & -N_{C}^{(0)\,*} & 0\\
-U_{12}^{T} & -U_{22}^{T} & 0 & 0\end{array}\right).\]
This block structure is still not beneficial for the simplification
of the corresponding pfaffian and we need to use the exchange matrices
defined in appendix \ref{sec:ExchMat} to exchange blocks 2 and 3
\[
\tilde{\mathbb{M}}_{R}=P_{23}\tilde{\mathbb{M}}P_{23}^{T}=\left(\begin{array}{cc|cc}
N_{C}^{(1)} & U_{11} & 0 & U_{12}\\
-U_{11}^{T} & -N_{C}^{(0)\,*} & -U_{21}^{T} & 0\\
\hline 0 & U_{21} & 0 & U_{22}\\
-U_{12}^{T} & 0 & -U_{22}^{T} & 0\end{array}\right).\]
Using the formulas of appendix \ref{sec:BipartitePfaff} for the pfaffian
of a bipartite matrix we obtain\begin{eqnarray*}
\textrm{Pf}(\tilde{\mathbb{M}}_{R}) & = & \textrm{Pf}\left(\begin{array}{cc}
0 & U_{22}\\
-U_{22}^{T} & 0\end{array}\right)\\
 & \times & \textrm{Pf}\left[\left(\begin{array}{cc}
N_{C}^{(1)} & X_{12}\\
-X_{12}^{T} & -N_{C}^{(0)\,*}\end{array}\right)\right]\end{eqnarray*}
where\begin{equation}
X_{12}=U_{11}+U_{12}U_{22}^{-1}U_{21}\label{eq:X12}\end{equation}
The first pfaffian in the right hand side of the above expression
is simply given by $(-1)^{K_{S}}\det U_{22}$ whereas the second pfaffian
can be computed using again the expression for the pfaffian of a bipartite
matrix. Collecting all the terms together we finally obtain\begin{eqnarray*}
\textrm{Pf}(\tilde{\mathbb{M}}_{R}) & = & (-1)^{K_{S}}\det(U_{22})\textrm{Pf}\left(N_{C}^{(1)}\right)\\
 & \times & \textrm{Pf}\left(-N_{C}^{(0)\,*}+X_{12}^{T}N_{C}^{(1)\,-1}X_{12}\right).\end{eqnarray*}
Taking into account that $\det(P_{23})=1$ we finally obtain\begin{eqnarray}
\textrm{pf}(\mathbb{M}) & = & (-1)^{K_{S}}\det(D_{11}^{(1)})\det(D_{11}^{(0)\,*})\det(U_{22})\textrm{Pf}\left(N_{C}^{(1)}\right)\nonumber \\
 & \times & \textrm{Pf}\left(-N_{C}^{(0)\,*}+X_{12}^{T}N_{C}^{(1)\,-1}X_{12}\right)\label{eq:pfMFinal}\end{eqnarray}

The advantage of this ugly result over the general expression is that
the dimensionality of the $N_{C}^{(1)}$, $N_{C}^{(0)}$, $D_{11}^{(i)}$
and $U_{11}$ matrices is $2(N-K_{S})$ which is much smaller than
the one of the original problem ($2N)$. The only big matrix in Eq.
(\ref{eq:X12}) is the inverse of $U_{22}$ with dimension $2K_{S}$,
however, this does not pose a challenge as its special structure Eq.
(\ref{eq:DefU22}) is very well adapted to the use of the Sherman-Morrison
formulas for the determinant and inverse of this kind of special matrices
\cite{NR.07}.

\subsection{Using a different reference vacuum\label{sub:DiffVac}}

From the above discussion, it is clear that the structure and properties
of the $M^{(i)}$ matrices is intimately related to the reference
vacuum used to express the HFB wave functions $|\phi_{i}\rangle$.
This suggests to use another reference vacuum, instead of the true
vacuum implicitly assumed in the previous discussions, with the hope
that the new matrices $\overline{M}^{(i)}$ will acquire a structure
where there would be no fully occupied quasiparticles and the number
of empty ones will be very large and comparable to the size of the
basis. Matrices with that properties will not require the use of the
{}``fully occupied limit'' formulas and will benefit from the reduction
in computational burden of the {}``fully empty limit'' results.
To be more specific, let us consider the example where the $|\phi_{i}\rangle$
correspond to two HFB wave functions with the same average of the
number of particles and different quadrupole deformation parameters
$q_{2}^{(i)}.$ It looks rather intuitive that a more appropriate
reference vacuum than the true vacuum could be the HFB wave function
$|\bar{\phi}\rangle$ with the same average of the number of particles
and a deformation parameter $\bar{q}_{2}$close to both $q_{2}^{(i)}$
(for instance the mean value $\frac{1}{2}(q_{2}^{(0)}+q_{2}^{(1)})$).
It is to be expected that, with respect to this reference vacuum $|\bar{\phi}\rangle$,
the new $\overline{U}^{(i)}$ amplitudes will be very close to the
identity matrix whereas the new $\overline{V}^{(i)}$ amplitudes will
be small. In other words, the HFB wave functions $|\phi_{i}\rangle$
will be represented by a linear combination of quasiparticle excitations
of the reference state $|\bar{\phi}\rangle$ with small amplitudes
that quickly decrease with the number of quasiparticle excitations
(i.e. the amplitudes of the four quasiparticle excitations much smaller
that the amplitudes of the two quasiparticle ones). The expected properties
of the new $\overline{U}^{(i)}$ and $\overline{V}^{(i)}$ amplitudes
imply that most of the quasiparticle excitations will correspond to
the {}``fully empty'' limit discussed in the previous subsection
(with the associated advantages of having to deal with matrices of
small size) and far from the problematic fully occupied limit of subsection
\ref{sub:Occ}. 

Let us consider the new reference vacuum $|\bar{\phi}\rangle$ with
the associated creation $\bar{\alpha}^{+}$and annihilation $\bar{\alpha}$
quasiparticle operators which are defined in terms of a single particle
basis of creation and annihilation operators by means of linear combinations
involving the $\bar{U}$ and $\bar{V}$ amplitudes\[
\left(\begin{array}{c}
\bar{\alpha}\\
\bar{\alpha}^{+}\end{array}\right)=\left(\begin{array}{cc}
\bar{U}^{+} & \bar{V}^{+}\\
\bar{V}^{T} & \bar{U}^{T}\end{array}\right)\left(\begin{array}{c}
a\\
a^{+}\end{array}\right)=\bar{W}^{+}\left(\begin{array}{c}
a\\
a^{+}\end{array}\right).\]
The same relation holds true for the quasiparticle operators $\alpha^{(i)}$
and $\alpha^{(i)\,+}$ with amplitudes $W^{(i)}$. Using the unitarity
of the matrices $W^{(i)}$ and $\bar{W}$ we can express the set of
quasiparticle operators $\alpha^{(i)}$ and $\alpha^{(i)\,+}$ in
terms of the $\bar{\alpha}$ and $\bar{\alpha}^{+}$ ones as\[
\left(\begin{array}{c}
\alpha^{(i)}\\
\alpha^{(i)\,+}\end{array}\right)=W^{(i)\,+}\bar{W}\left(\begin{array}{c}
\bar{\alpha}\\
\bar{\alpha}^{+}\end{array}\right)=\bar{W}^{(i)\,+}\left(\begin{array}{c}
\bar{\alpha}\\
\bar{\alpha}^{+}\end{array}\right)\]
Using Thouless theorem we can also express the $|\bar{\phi}_{i}\rangle$
wave functions ( satisfying $\langle\bar{\phi}|\bar{\phi}^{(i)}\rangle=1$
and therefore different from the previous $|\phi_{i}\rangle$ by a
normalization factor) in terms of the $|\bar{\phi}\rangle$ reference
vacuum \begin{equation}
|\bar{\phi}_{i}\rangle=\exp\left(\frac{1}{2}\sum_{kk'}\bar{M}_{kk'}^{(i)}\bar{\alpha}_{k}^{+}\bar{\alpha}_{k'}^{+}\right)|\bar{\phi}\rangle.\label{eq:HFBWFBAR}\end{equation}
with $\bar{M}^{(i)}=(\bar{V}^{(i)}\bar{U}^{(i)\,-1})^{*}$ (not to
be confused with the diagonal matrix of Eq (\ref{eq:Mbarave})). These
two matrices can be easily computed once the $\bar{W}$ coefficients
have been given and it is even possible to give an analytical expression
\cite{Blaizot.81} in terms of $\bar{M}$ and $M^{(i)}$\begin{equation}
\bar{M}^{(i)}=Q(\bar{M}^{T}+M^{(i)})(\mathbb{I}+\bar{M}^{+}M^{(i)})^{-1}(Q^{+})^{-1}.\label{eq:Mbarave}\end{equation}
with $Q=(\mathbb{I}+\bar{M}\bar{M}^{+})^{-1/2}$. All the formulas
given above (and below) are equally valid for the wave functions given
in the form of Eq (\ref{eq:HFBWFBAR}) with the amplitudes of Eq.
(\ref{eq:Mbarave}).

\section{Different single particle bases\label{sec:DiferentBases}}

It is very common that the HFB wave functions $|\phi_{i}\rangle$
are defined in terms of different single particle basis, with creation
and annihilation operators $a_{k}^{+}(i)$ and $a_{k}(i)$ that will
carry indexes $(i)$ to indicate the HFB wave function they belong
to. Those bases are usually not complete and therefore they span different
subspaces of the full Hilbert space. As a consequence, the formulas
obtained above cannot be used because they implicitly rely on a common
basis for the two HFB wave functions \cite{Robledo.94}. The strategy
to overcome this problem is to find a bigger subspace encompassing
both subspace and use an orthogonal basis defined in that bigger subspace
for the two HFB wave functions $|\phi_{i}\rangle$. In an early consideration
of this problem \cite{Robledo.94} I used the whole Hilbert space
as the common subspace. Another possibility, explored in this paper,
is to consider the subspace generated by the union of the two subspaces.
In this case, special care has to be taken with the resulting basis,
union of the of the two original bases, as it can be redundant (i.e.
it can contain linearly dependent vectors). 

Let me start by considering the two basis $\{a_{k}^{+}(0),\, k=0,\ldots,N_{(0)}\}$
and $\{a_{k}^{+}(1),\, k=0,\ldots,N_{(1)}\}$ which are defined in
terms of single particle creation operators (typically those of the
harmonic oscillator basis) and with dimensions $N_{(0)}$ and $N_{(1)}$,
respectively, which do not necessarily coincide. It is worth introducing
the set of creation operators\[
A_{\mu}^{+}=\begin{cases}
a_{k}^{+}(0) & \mu=k\,\, k=1,\ldots,N_{(0)}\\
a_{l}^{+}(1) & \mu=l+N_{(0)}\,\, l=1,\ldots,N_{(1)}\end{cases}\]
embracing the two sets of creation operators of the bases. They satisfy
the commutation relations $\{A_{\mu},\, A_{\nu}^{+}\}=\mathcal{N}_{\mu\nu}$
and $\{A_{\mu},\, A_{\nu}\}=\{A_{\mu}^{+},\, A_{\nu}^{+}\}=0$ where
the overlap matrix $\mathcal{N}$ (dimension $(N_{(0)}+N_{(1)})\times(N_{(0)}+N_{(1)})$
is given in terms of the rectangular matrix $T_{k\, k'}=\{a_{k}(0),a_{k'}^{+}(1)\}=\langle-|a_{k}(0)a_{k'}^{+}(1)|-\rangle$
by the expression\begin{equation}
\mathcal{N}=\left(\begin{array}{cc}
\mathbb{I}_{(0)} & T\\
T^{+} & \mathbb{I}_{(1)}\end{array}\right).\label{eq:DefN}\end{equation}
The overlap matrix is hermitian, semi-positive definite and therefore
can be diagonalized by a unitary transformation $D$, i.e. $\mathcal{N}=DnD^{+}$where
the diagonal matrix $n$ of the eigenvalues is of dimension $(N_{(0)}+N_{(1)})\times(N_{(0)}+N_{(1)})$.
In order to deal with the zero (or smaller than a given threshold)
eigenvalues case (which signals the appearance of linearly dependent
basis states) it is convenient to introduce the notation \[
n=\left(\begin{array}{cc}
\bar{n} & 0\\
0 & \epsilon\end{array}\right).\]
where $\bar{n}$ is a diagonal matrix with the eigenvalues different
from zero and $\epsilon$ is the diagonal matrix of dimension $N_{\epsilon}$containing
those eigenvalues with value zero (or smaller than a numerical threshold).
It is convenient in the ensuing developments to consider that the
matrix $\epsilon$ is different from zero and therefore can be inverted.
At the end of the calculations $\epsilon$ will be made to tend to
zero to obtain the final result. Taking this regularization scheme
into account, we can define the square root of the overlap matrix
$\mathcal{N}^{1/2}=Dn^{1/2}D^{+}$ and its inverse $\mathcal{N}^{-1/2}=Dn^{-1/2}D^{+}$
that are required to define the operators \[
B_{\mu}=\sum_{\nu}\mathcal{N}_{\mu\nu}^{-1/2}A_{\nu}\]
as well as the inverse relation $A_{\mu}=\sum_{\nu}\mathcal{N}_{\mu\nu}^{1/2}B_{\nu}$.
The creation and annihilation operators $B_{\mu}^{+}$ and $B_{\nu}$
satisfy canonical commutation relations $\{B_{\mu},B_{\nu}^{+}\}=\left(\mathcal{N}^{-1/2}\mathcal{N}\mathcal{N}^{-1/2}\right)_{\mu\nu}=\delta_{\mu\nu}$.
They are introduced to express the HFB wave functions of Eq. (\ref{eq:HFBWFBAR})
in the standard way as\[
|\phi_{i}\rangle=\exp\left\{ \sum_{\mu\mu'}\frac{1}{2}\tilde{N}_{\mu\mu'}^{(i)}B_{\mu}^{+}B_{\mu'}^{+}\right\} |0\rangle\]
with the matrices of dimension $(N_{(0)}+N_{(1)})\times(N_{(0)}+N_{(1)})$
\begin{equation}
\tilde{N}^{(i)}=\mathcal{N}^{1/2\,+}\tilde{M}^{(i)}\mathcal{N}^{1/2\,*}\label{eq:Ntilda}\end{equation}
 given in terms of the extended matrices (also of dimension $(N_{(0)}+N_{(1)})\times(N_{(0)}+N_{(1)})$)
\begin{equation}
\tilde{M}^{(0)}=\left(\begin{array}{cc}
M^{(0)} & 0\\
0 & 0\end{array}\right),\:\tilde{M}^{(1)}=\left(\begin{array}{cc}
0 & 0\\
0 & M^{(1)}\end{array}\right).\label{eq:MtildaExt}\end{equation}
As the operators $B_{\mu}^{+}$ and $B_{\nu}$ satisfy canonical commutation
relations, it is now possible to apply the standard formalism already
developed in Ref \cite{Robledo.09} to write \[
\langle\phi_{0}|\phi_{1}\rangle=S_{N_{(0)}+N_{(1)}}\textrm{pf}(\tilde{\mathbb{M})}\]
where the matrix $\tilde{\mathbb{M}}$ entering the argument of the
pfaffian is given in terms of the matrices defined in Eq (\ref{eq:MtildaExt})
as\[
\tilde{\mathbb{M}}=\left(\begin{array}{cc}
\tilde{N}^{(1)} & -\mathbb{I}\\
\mathbb{I} & -\tilde{N}^{(0)\,*}\end{array}\right).\]
Using the results of appendix \ref{sec:Overlap} concerning the eigenvalues
and eigenvectors of the norm matrix $\mathcal{N}$ (to simplify the
notation in the following we consider the size of the two basis to
be equal $N_{(0)}=N_{(1)}=N$) and using the definition of Eq (\ref{eq:Ntilda})
we have \begin{eqnarray*}
\tilde{N}^{(i)} & = & D\left(\begin{array}{cc}
n_{+}^{1/2} & 0\\
0 & n_{-}^{1/2}\end{array}\right)D^{+}\tilde{M}^{(i)}D^{*}\left(\begin{array}{cc}
n_{+}^{1/2} & 0\\
0 & n_{-}^{1/2}\end{array}\right)D^{T}\\
 & = & D\tilde{N}_{D}^{(i)}D^{T}\end{eqnarray*}
which defines the matrices $\tilde{N}_{D}^{(i)}$ as \[
\tilde{N}_{D}^{(i)}=\left(\begin{array}{cc}
n_{+}^{1/2} & 0\\
0 & n_{-}^{1/2}\end{array}\right)D^{+}\tilde{M}^{(i)}D^{*}\left(\begin{array}{cc}
n_{+}^{1/2} & 0\\
0 & n_{-}^{1/2}\end{array}\right).\]
Please note that these matrices are well defined when some of the
eigenvalues of the norm overlap matrix $\mathcal{N}$ go to zero (i.e.
some of the elements of the diagonal matrix $n_{.}$ are zero). Using
the explicit form of the matrix $D$ given in appendix \ref{sec:Overlap}
in terms of the matrices $E$ and $F$ entering the Singular Value
Decomposition (SVD) of the matrix $T$ and defining the auxiliary
matrices $E_{\pm}=n_{\pm}^{1/2}E$ and $F_{\pm}=n_{\pm}^{1/2}F$ we
get\[
\tilde{N}_{D}^{(0)}=\frac{1}{2}\left(\begin{array}{cc}
E_{+} & 0\\
0 & E_{-}\end{array}\right)\left(\begin{array}{cc}
M^{(0)} & -M^{(0)}\\
-M^{(0)} & M^{(0)}\end{array}\right)\left(\begin{array}{cc}
E_{+}^{T} & 0\\
0 & E_{-}^{T}\end{array}\right)\]
and\[
\tilde{N}_{D}^{(1)}=\frac{1}{2}\left(\begin{array}{cc}
F_{+} & 0\\
0 & F_{-}\end{array}\right)\left(\begin{array}{cc}
M^{(1)} & M^{(1)}\\
M^{(1)} & M^{(1)}\end{array}\right)\left(\begin{array}{cc}
F_{+}^{T} & 0\\
0 & F_{-}^{T}\end{array}\right)\]
Finally, using known properties of the pfaffian we can express the
overlap in terms of the $\tilde{N}_{D}^{(i)}$ as\begin{equation}
\langle\phi_{0}|\phi_{1}\rangle=(-1)^{N}\textrm{pf}\left(\begin{array}{cc}
\tilde{N}_{D}^{(1)} & -\mathbb{I}\\
\mathbb{I} & -\tilde{N}_{D}^{(0)\,*}\end{array}\right)\label{eq:Overlap_F}\end{equation}
which is the final expression for the overlap.

To finish this section it is worth considering the limit where the
two bases are connected by means of an unitary transformation (see
appendix \ref{sec:Overlap}). In this case, the SVD of $T$ is trivial
and we have $n_{+}^{1/2}=\sqrt{2}\mathbb{I}$, $n_{-}^{1/2}=0$, $E=\mathbb{I}$
and $F=T$. Using these values we have\[
\tilde{N}_{D}^{(1)}=\left(\begin{array}{cc}
TM^{(1)}T^{T} & 0\\
0 & 0\end{array}\right)\]
and \[
\tilde{N}_{D}^{(0)}=\left(\begin{array}{cc}
M^{(0)} & 0\\
0 & 0\end{array}\right).\]
As a consequence of the zero eigenvalues of the overlap matrix, the
matrices $\tilde{N}_{D}^{(i)}$ acquire a bipartite structure where
only the upper diagonal block is different from zero. In this case,
the matrix in the pfaffian in Eq. (\ref{eq:Overlap_F}) becomes a
block matrix \[
\left(\begin{array}{cc|cc}
TM^{(1)}T^{T} & 0 & -\mathbb{I} & 0\\
0 & 0 & 0 & -\mathbb{I}\\
\hline \mathbb{I} & 0 & -M^{(0)\,*} & 0\\
0 & \mathbb{I} & 0 & 0\end{array}\right).\]
The pfaffian of this matrix can be simplified by using the results
of appendix \ref{sec:ExchMat} and, after exchanging blocks 2 and
3, bring the matrix to block diagonal form. Once in block diagonal
form, the pfaffian can be reduced to the product of the pfaffian of
each of the diagonal blocks by using a simplified version of the results
of appendix \ref{sec:BipartitePfaff} for the pfaffian of a bipartite
matrix. The final results is  \[
\langle\phi_{0}|\phi_{1}\rangle=(-1)^{N(N+1)/2}\textrm{pf}\left(\begin{array}{cc}
TM^{(1)}T^{T} & -\mathbb{I}\\
\mathbb{I} & -M^{(0)\,*}\end{array}\right)\]
as expected (see Ref \cite{Robledo.09}).

\section{Conclusions}

We have analyzed the formula for the evaluation of the norm overlap
of two different HFB wave functions in the situation where some of
the occupancies of the quasiparticle levels are one and the standard
approach leads to indeterminacies what have to be singled out in order
to obtain a meaningful answer. In the case where there are fully empty
single particle levels the overlap formula is well behaved but this
limit is also addressed as a (significant) reduction in the computational
burden can be obtained if the number of empty levels is large enough.
Finally, the case where each of the two HFB wave functions are expressed
in different single particle basis is addressed and the formalism
to compute the overlap in this situation is developed. A common basis,
union of the other two is considered, and special attention has to
be paid to the problem of redundancy of the enlarged subspace. The
formulas given in this paper are a practical complement of the general
one given in \cite{Robledo.09} and should be useful for a practical
implementation of the calculation of the overlaps of two different
HFB wave functions in the most general case.
\begin{acknowledgments}
\appendix
This work was supported by MICINN (Spain) under research grants No.
FPA2009-08958, and No. FIS2009-07277, as well as by Consolider-Ingenio
2010 Programs CPAN CSD2007-00042 and MULTIDARK CSD2009-00064.
\end{acknowledgments}

\section{Reordering of matrices and its impact in the pfaffian\label{sec:ExchMat}}

In some situations we will have to make use of eventual peculiarities
of the block structure of the skew-symmetric matrix in order to simplify
the final expression of the pfaffian. To do so, it is convenient to
know how to reorder rows and columns of a matrix as well as the impact
of such reordering in the pfaffian. A useful set of matrices is the
one of the matrices $E(i,j)$ that exchanges columns $i$ and $j$
of a matrix when multiplied to the right hand side of that matrix,
i.e.\begin{eqnarray*}
\left(\begin{array}{cccclcccc}
\ldots & a_{i-1} & a_{i} & a_{i+1} & \ldots & a_{j-1} & a_{j} & a_{j+1} & \ldots\\
\vdots & \vdots & \vdots & \vdots & \vdots & \vdots & \vdots & \vdots & \vdots\end{array}\right)E(i,j) & =\\
\left(\begin{array}{clccccccc}
\ldots & a_{i-1} & a_{j} & a_{i+1} & \ldots & a_{j-1} & a_{i} & a_{j+1} & \ldots\\
\vdots & \vdots & \vdots & \vdots & \vdots & \vdots & \vdots & \vdots & \vdots\end{array}\right)\end{eqnarray*}
The matrices$E(ij)$ are characterized by the matrix elements\[
E(i,j)_{kl}=\delta_{kl}-\delta_{ki}\delta_{li}-\delta_{kj}\delta_{lj}+\delta_{kj}\delta_{li}+\delta_{ki}\delta_{lj}\]
and are unit matrices where the elements $i$ and $j$ of the diagonal
are set to zero and the elements $i,j$ and $j,i$ are set to one. 

Another useful set of matrices is the one of the $S(i,j)$ matrices
such that, when multiplied to the right hand side of a matrix, moves
the column $j$ of the matrix to the position of column $i$ ($i<j$)
and then shifts column $i$ to position $i+1,$ column $i+1$ to position
$i+2$, and so on, up to column $j-1$ that is shifted to column $j$,
i.e. \begin{eqnarray*}
\left(\begin{array}{cccccccccc}
\ldots & a_{i-1} & a_{i} & a_{i+1} & a_{i+2} & \ldots & a_{j-1} & a_{j} & a_{j+1} & \ldots\\
\vdots & \vdots & \vdots & \vdots & \vdots & \vdots & \vdots & \vdots & \vdots & \vdots\end{array}\right)S(i,j) & =\\
\left(\begin{array}{cccccccccc}
\ldots & a_{i-1} & a_{j} & a_{i} & a_{i+1} & a_{i+2} & \ldots & a_{j-1} & a_{j+1} & \ldots\\
\vdots & \vdots & \vdots & \vdots & \vdots & \vdots & \vdots & \vdots & \vdots & \vdots\end{array}\right)\end{eqnarray*}
In terms of matrix elements they are given by \[
S(i,j)_{kl}=\delta_{kl}-\sum_{s=i}^{j}\delta_{ks}\delta_{ls}+\sum_{s=i}^{j-1}\delta_{ks}\delta_{ls+1}+\delta_{kj}\delta_{li}.\]
These matrices are unit matrices where the $1$ in position $i,i$
is shifted to position $i,i+1$ , the $1$ in position $i+1,i+1$
is shifted to position $i+1,i+2$ and so on up the $1$ in position
$j,j$ that is shifted to position $j,i$. The determinants of the
two kind of matrices are easy to determine and they are given by \[
\det(E(i,j))=-1\]
 and \[
\det(S(i,j))=(-1)^{j-i}.\]
The successive application of the matrices $E(i+k,j+k)$ for $k=0$
up to $k=N$ defines a matrix \[
P_{N}(i,j)=\prod_{k=0}^{N-1}E(i+k,j+k)\]
that exchanges a set of $N$ columns at once\begin{eqnarray*}
\left(\begin{array}{ccccccccc}
\ldots & a_{i} & \ldots & a_{i+N-1} & \ldots & a_{j} & \ldots & a_{j+N-1} & \ldots\\
\vdots & \vdots & \vdots & \vdots & \vdots & \vdots & \vdots & \vdots & \vdots\end{array}\right)P_{N}(i,j) & =\\
\left(\begin{array}{ccccccccc}
\ldots & a_{j} & \ldots & a_{j+N-1} & \ldots & a_{i} & \ldots & a_{i+N-1} & \ldots\\
\vdots & \vdots & \vdots & \vdots & \vdots & \vdots & \vdots & \vdots & \vdots\end{array}\right)\end{eqnarray*}
Applying the matrix $P_{N}^{T}$ to the left of the matrix the corresponding
exchange of rows is produced. As a consequence \begin{eqnarray}
P_{N}^{T}(i,j)\left(\begin{array}{ccccc}
\ddots & \vdots & \vdots & \vdots\\
\cdots & A_{N}(i,i) & \cdots & A_{N}(i,j) & \cdots\\
\vdots & \vdots & \ddots & \vdots & \vdots\\
\cdots & A_{N}(j,i) & \cdots & A_{N}(j,j) & \cdots\\
 & \vdots & \vdots & \vdots & \ddots\end{array}\right)P_{N}(i,j) & =\nonumber \\
\left(\begin{array}{ccccc}
\ddots & \vdots & \vdots & \vdots\\
\cdots & A_{N}(j,j) & \cdots & A_{N}(j,i) & \cdots\\
\vdots & \vdots & \ddots & \vdots & \vdots\\
\cdots & A_{N}(i,j) & \cdots & A_{N}(i,i) & \cdots\\
 & \vdots & \vdots & \vdots & \ddots\end{array}\right)\label{eq:PTAP}\end{eqnarray}
where $A_{N}(i,j)$ are sub-matrices of dimension $N\times N$ whose
first element is located in the row $i$ and column $j$ of the matrix
where the $A_{N}$ are embedded. The result obtained, together with
that of the pfaffian of a bipartite matrix will be useful to reduce
some of the pfaffians encountered in the main body of the paper. Obviously
$\det P_{N}=(-1)^{N}$. Unfortunately, this trick can not be applied
when the number of columns to be {}``exchanged'' is not the same.
In such case, we have to consider the more general shift operation
$S(i,j)$. To see how it works let us consider a matrix $A$ were
there are three groups of columns denoted by $L$, $C,$ and $R$
such that the first group goes from column $i$ to columns $i+N_{L}-1$
(i.e. $N_{L}$ columns), the group $C$ goes from column $i+N_{L}$
up to columns $i+N_{L}+N_{C}-1$ (i.e. $N_{C}$ columns) and finally
the group $R$ from column $i+N_{L}+N_{C}$ up to column $i+N_{L}+N_{C}+N_{R}-1$
(i.e. $N_{R}$ columns). Schematically, the columns of the matrix
$A$ could be represented as\[
A=\left(\ldots|L|C|R|\ldots\right)\]
Now we want to exchange the group $L$ with the group $R$; to do
so the group $R$ of columns is moved to the position occupied by
$L$ using the product of matrices \[
P_{LR}^{(1)}=\prod_{k=0}^{N_{R}-1}S(i+k,i+k+N_{L}+N_{C})\]
 giving \[
\left(\ldots|L|C|R|\ldots\right)P_{LR}^{(1)}=\left(\ldots|R|L|C|\ldots\right)\]
Now, the group of columns $C$ is moved to the position of the group
$L$ by means of the following product of {}``shift'' matrices \[
P_{LC}=\prod_{k=0}^{N_{C}-1}S(i+k+N_{R},i+k+N_{R}+N_{L})\]
Using this matrix we obtain\begin{eqnarray*}
\left(\ldots|L|C|R|\ldots\right)P_{LR}^{(1)}P_{LC} & = & \left(\ldots|R|L|C|\ldots\right)P_{LC}\\
 & = & \left(\ldots|R|C|L|\ldots\right)\end{eqnarray*}
The matrix exchanging the set of columns $L$ with the set $R$ will
be denoted $P_{LR}=P_{LR}^{(1)}P_{LC}.$ By applying $P_{LR}^{T}$
to the left of the matrix the set of rows $L$ and $R$ are exchanged.
As a consequence \begin{eqnarray*}
P_{LR}^{T}(i,j)\left(\begin{array}{ccccc}
\ddots & \vdots & \vdots & \vdots\\
\cdots & A_{N_{L}\times N_{L}}(i,i) & \cdots & A_{N_{L}\times N_{R}}(i,j) & \cdots\\
\vdots & \vdots & \ddots & \vdots & \vdots\\
\cdots & A_{N_{R}\times N_{L}}(j,i) & \cdots & A_{N_{R}\times N_{R}}(j,j) & \cdots\\
 & \vdots & \vdots & \vdots & \ddots\end{array}\right)P_{LR}(i,j) & =\\
\left(\begin{array}{clccc}
\ddots & \vdots & \vdots & \vdots\\
\cdots & A_{N_{R}\times N_{R}}(j,j) & \cdots & A_{N_{R}\times N_{L}}(j,i) & \cdots\\
\vdots & \vdots & \ddots & \vdots & \vdots\\
\cdots & A_{N_{L}\times N_{R}}(i,j) & \cdots & A_{N_{L}\times N_{L}}(i,i) & \cdots\\
 & \vdots & \vdots & \vdots & \ddots\end{array}\right)\end{eqnarray*}
where $A_{N_{L}\times N_{L}}(i,i)$ is a sub-matrix with $N_{L}$
rows and $N_{L}$ columns starting at row $i$ and column $i$, $A_{N_{L}\times N_{R}}(i,j)$
is a sub-matrix with $N_{L}$ rows and $N_{R}$ columns starting at
row $i$ and column $j=i+N_{L}+N_{C}$, and so on. This result generalizes
the one of Eq (\ref{eq:PTAP}) for sub-matrices of different sizes.
The impact of such exchange of rows and columns in the pfaffian is
the product of the determinants of the $S$ matrices involved, i.e.
$\det(P_{LR})=\det(P_{LR}^{(1)})\det(P_{LC}$). With $\det(P_{LR}^{(1)})=(-1)^{(N_{L}+N_{C})N_{R}}$
and $\det(P_{LC})=(-1)^{N_{L}N_{C}}$ the total phase is $\det(P_{LR})=(-1)^{N_{L}N_{R}+N_{C}N_{R}+N_{L}N_{C}}$
that reduces to $(-1)^{N}$ when $N_{L}=N_{R}=N$, a result that is
independent of $N_{C}$.

\section{The pfaffian of a bipartite matrix\label{sec:BipartitePfaff}}

To derive some of the results obtained in this paper, it is often
required to compute the pfaffian of a bipartite skew-symmetric matrix
with the general structure\begin{equation}
S=\left(\begin{array}{cc}
M & Q\\
-Q^{T} & N\end{array}\right)\label{eq:BP_S}\end{equation}
where $M$ and $N$ are skew-symmetric matrices and $Q$ is a general
rectangular matrix. Using Aitken's formula it is possible to diagonalize
the bipartite matrix using a congruence transformation \begin{eqnarray*}
\left(\begin{array}{cc}
\mathbb{I} & 0\\
Q^{T}M^{-1} & \mathbb{I}\end{array}\right)\left(\begin{array}{cc}
M & Q\\
-Q^{T} & N\end{array}\right)\left(\begin{array}{cc}
\mathbb{I} & -M^{-1}Q\\
0 & \mathbb{I}\end{array}\right) & =\\
\left(\begin{array}{cc}
M & 0\\
0 & N+Q^{T}M^{-1}Q\end{array}\right)\end{eqnarray*}
or the equivalent expression \begin{eqnarray*}
\left(\begin{array}{cc}
\mathbb{I} & -QN^{-1}\\
0 & \mathbb{I}\end{array}\right)\left(\begin{array}{cc}
M & Q\\
-Q^{T} & N\end{array}\right)\left(\begin{array}{cc}
\mathbb{I} & 0\\
N^{-1}Q^{T} & \mathbb{I}\end{array}\right) & =\\
\left(\begin{array}{cc}
M+QN^{-1}Q^{T} & 0\\
0 & N\end{array}\right)\end{eqnarray*}
to be used if $M^{-1}$ does not exist. This is a very convenient
block diagonalization formula as it involves congruence transformations
that allow to use the property $\textrm{pf}(P^{T}RP)=\textrm{det}(P)\textrm{pf}(R)$
of the pfaffian to obtain the next two identities\begin{eqnarray}
\textrm{pf}(S) & = & \textrm{pf}(M)\textrm{pf}(N+Q^{T}M^{-1}Q)\label{eq:BP_PfS}\\
 & = & \textrm{pf}(M+QN^{-1}Q^{T})\textrm{pf}(N)\end{eqnarray}

\section{The overlap matrix\label{sec:Overlap}}

The overlap matrix $\mathcal{N}$ has the bipartite structure\[
\mathcal{N}=\left(\begin{array}{cc}
\mathbb{I}_{(0)} & T\\
T^{+} & \mathbb{I}_{(1)}\end{array}\right)\]
where the rectangular matrix $T$, with matrix elements $T_{k\, k'}=\{a_{k}(0),a_{k'}^{+}(1)\}=\langle0|a_{k}(0)a_{k'}^{+}(1)|0\rangle$,
is the matrix of the overlaps between the elements of the two basis
considered. As in the body of the paper, $N_{(0)}$ and $N_{(1)}$
denote the dimensions of the each of the basis and it is assumed for
definiteness that $N_{(0)}\geq N_{(1)}$.The matrices $\mathbb{I}_{(0)}$
and $\mathbb{I}_{(1)}$ stand for the identity matrices of dimensions
$N_{(0)}$ and $N_{(1)}$, respectively. For the developments considered
in this paper, the analysis of the spectral decomposition of the overlap
matrix is required in order to handle properly the occurrence of very
small or zero eigenvalues of the overlap. The analysis is based on
the Singular Value Decomposition (SVD) \cite{Golub.96} of the matrix
$T$ \begin{equation}
T=E^{+}\Delta F\label{eq:SVD_T}\end{equation}
where $E$ and $F$ are square unitary matrices of dimensions $N_{(0)}\times N_{(0)}$
and $N_{(1)}\times N_{(1)}$ respectively and $\Delta$ is a rectangular
matrix of dimension $N_{(0)}\times N_{(1)}$ with the {}``diagonal
structure''\[
\Delta=\left(\begin{array}{c}
\bar{\Delta}\\
0\end{array}\right)\]
where $\bar{\Delta}$ is a real and positive square diagonal matrix
with dimension $N_{(1)}\times N_{(1)}$. It is convenient to introduce
a rectangular {}``identity matrix'' $\mathbb{I}_{(01)}$ of dimension
$N_{(0)}\times N_{(1)}$ with a structure similar to the matrix $\Delta$,
namely \[
\mathbb{I}_{(01)}=\left(\begin{array}{c}
\mathbb{I}_{(1)}\\
0\end{array}\right)\]
and such that $\Delta=\mathbb{I}_{(01)}\bar{\Delta}$. This matrix
also has the property $\mathbb{I}_{(01)}^{+}\mathbb{I}_{(01)}=\mathbb{I}_{(1)}$.

Using the SVD of $T$ defined in Eq. (\ref{eq:SVD_T}) we can finally
write\[
\mathcal{N}=\bar{D}\left(\begin{array}{cc}
\mathbb{I}_{(0)} & \Delta\\
\Delta^{+} & \mathbb{I}_{(1)}\end{array}\right)\bar{D}^{+}\]
where\[
\bar{D}=\left(\begin{array}{cc}
E^{+} & 0\\
0 & F^{+}\end{array}\right).\]
The matrix in the middle can be easily brought to diagonal form \begin{eqnarray*}
 & \bar{D}^{(0)\,+}\left(\begin{array}{cc}
\mathbb{I}_{(0)} & \Delta\\
\Delta^{+} & \mathbb{I}_{(1)}\end{array}\right)\bar{D}^{(0)}=\\
 & \left(\begin{array}{cc}
2\mathbb{I}_{(0)}+\mathbb{I}_{(01)}(\bar{\Delta}-\mathbb{I}_{(1)})\mathbb{I}_{(01)}^{+} & 0\\
0 & \mathbb{I}_{(1)}-\bar{\Delta}\end{array}\right)\end{eqnarray*}
by means of the $\bar{D}^{(0)}$ transformation\[
\bar{D}^{(0)}=\frac{1}{\sqrt{2}}\left(\begin{array}{cc}
\mathbb{I}_{(0)} & -\mathbb{I}_{(01)}\\
\mathbb{I}_{(01)}^{+} & \mathbb{I}_{(1)}\end{array}\right)\]
Finally, introducing the matrix $D=\bar{D}\bar{D}^{(0)}$ we obtain
the complete diagonalization of the overlap matrix \begin{equation}
\mathcal{N}=D\left(\begin{array}{cc}
2\mathbb{I}_{(0)}+\mathbb{I}_{(01)}(\bar{\Delta}-\mathbb{I}_{(1)})\mathbb{I}_{(01)}^{+} & 0\\
0 & \mathbb{I}_{(1)}-\bar{\Delta}\end{array}\right)D^{+}.\label{eq:NDiagG}\end{equation}
The semi-positive character of the matrix $\mathcal{N}$ implies that
all its eigenvalues are positive or zero and therefore (lower block)
the elements of the diagonal matrix $\bar{\Delta}$ cannot exceed
one. In the case of bases with equal dimensions $N_{(0)}=N_{(1)}$
the above result becomes \begin{equation}
\mathcal{N}=D\left(\begin{array}{cc}
\mathbb{I}+\bar{\Delta} & 0\\
0 & \mathbb{I}-\bar{\Delta}\end{array}\right)D^{+}=D\left(\begin{array}{cc}
n_{+} & 0\\
0 & n_{-}\end{array}\right)D^{+}\label{eq:NDiagES}\end{equation}
where the diagonal matrices $n_{\pm}=\mathbb{I}\pm\bar{\Delta}$ have
been introduced. As the matrix $\bar{\Delta}$ is also positive semidefinite
it is clear that the zero norm eigenvalues of $\mathcal{N}$ are associated
to values of $\bar{\Delta}$ equal to one. 

It is helpful to consider the special case when $T$ is an unitary
matrix. It corresponds to the common situation where the two bases
are connected through a unitary transformation as, for instance, the
ones associated to symmetry operations like rotations in real space,
etc acting on closed basis. For unitary $T$, the SVD factors can
be chosen as $E^{+}=T$, $\Delta=\mathbb{I}$ and $F=\mathbb{I}$
in the formulas above so that \begin{equation}
D=\frac{1}{\sqrt{2}}\left(\begin{array}{cc}
T & -T\\
\mathbb{I} & \mathbb{I}\end{array}\right)\label{eq:DefDTunit}\end{equation}
and \begin{equation}
\mathcal{N}=D\left(\begin{array}{cc}
2\mathbb{I} & 0\\
0 & 0\end{array}\right)D^{+}\label{eq:NDiagTUnitary}\end{equation}
with $N$-fold degenerate eigenvalues 2 and 0. 

To finish this appendix, just mention that a similar (but less general)
treatment of the norm overlap was considered in Ref \cite{Berger.80}
but without resorting to the powerful concept of the SVD.

\end{document}